\documentclass[aps,preprint,groupeaddress,showpacs]{revtex4}

\usepackage{graphicx}% Include figure files

\newcommand{\bm}{\boldmath}
\newcommand{\bv}[1]{\mbox{\bm $#1$}}

\begin{document}

\title{Position and velocity space diffusion of test particles
in stochastic electromagnetic fields}
%\shorttitle{Particle diffusion in stochastic electromagnetic fields} %Insert here a short version of the title if it exceeds 70 characters
\author{Silvia Perri}
\affiliation{Dipartimento di Fisica, Universit\`a della Calabria,
Via P. Bucci 31/C, I-87036 Rende (CS), Italy}
\author{Fabio Lepreti} 
\email[Corresponding author. Email address: ]{lepreti@fis.unical.it}
\author{Vincenzo Carbone}
\affiliation{Dipartimento di Fisica, Universit\`a della Calabria,
Via P. Bucci 31/C, I-87036 Rende (CS), Italy}
\affiliation{Consorzio Nazionale Interuniversitario per le Scienze
Fisiche della Materia (CNISM),~Unit\`a~di~Cosenza, Italy}
\author{Angelo Vulpiani}
%\thanks{}
\affiliation{Dipartimento di Fisica and INFN, Universit\`a di Roma ``La Sapienza'' - Piazzale A. Moro 2, I-00185 Roma, Italy}

\date{\today}

\begin{abstract}
The two--dimensional diffusive dynamics of test particles in a random electromagnetic field is studied. The synthetic electromagnetic fluctuations are generated through randomly placed magnetised ``clouds'' oscillating with a frequency $\omega$. We investigate the mean square displacements of particles in both position and velocity spaces. As $\omega$ increases the particles undergo standard (Brownian--like) motion, anomalous diffusion and ballistic motion in position space. Although in general the diffusion properties in velocity space are not trivially related to those in position space, we find that energization is present only when particles display anomalous diffusion in position space. The anomalous character of the diffusion is only in the non--standard values of the scaling exponents while the process is Gaussian.
\end{abstract}


\maketitle

\section{Introduction}
Since the paper by Fermi on the origin of cosmic radiation \cite{fermi}, stochastic acceleration of charged particles in random electromagnetic fields is a topic of primary interest because it plays a crucial role in understanding a lot of dynamical processes in astrophysics, space physics and laboratory plasmas (among the huge literature cf. e.g. Ref.s \cite{libro,brandford,jokipii,miller,matthaeus,vlahos} and references therein). In the non--relativistic case stochastic acceleration can be viewed as a diffusion process in
position and velocity space and it can be studied using a dynamical system approach \cite{sistemi,papachristou,bouchet,NOI,karlis}. In this framework, the diffusive motion of particles in turbulent magnetic fields reveals a rich class of nonlinear phenomena, but the influence of electric fluctuations is not well understood, at least not at the same level as for magnetic fluctuations. When electric fluctuations are present, the electromagnetic field is not stationary, diffusion in velocity space can occur also in 2--D and it is at least as interesting as the usual diffusion in position space.
In particular the well known different diffusive regimes in position space, namely standard (i.e. brownian--like) and anomalous diffusion, are related in a nontrivial way to diffusive regimes in velocity space \cite{bouchet,stawicki}. In the present paper we investigate the diffusive dynamics of test particles subjected to a stochastic electromagnetic field. Electric and magnetic fluctuations are obtained through a random distribution of oscillating magnetised ``clouds''. Different diffusive regimes are found as the oscillation frequency of the clouds changes, and a nontrivial relationship between diffusive properties in position and velocity spaces is observed.

%%%%%%%%%%%%%%%%%%%%%%%%%%%%%%%%%%%%%%%%%%%%%%%%%%%%%%%%%%%%%%%%%%%%%%%

\section{The stochastic electromagnetic field model}

We study the non--relativistic behavior of charged particles interacting with a stochastic electromagnetic field, the motion of each test particle being described by the usual equations of motion 

\begin{equation}
\frac{d \bv{r}}{dt} = \bv{v} \, ,
\end{equation}
\begin{equation}
\frac{d \bv{v}}{d t} = \frac{q}{m} 
\left[ \bv{E}(\bv{r},t) + \bv{v} \times \bv{B}(\bv{r},t) \right] \; .
\label{eq-v}
\end{equation}
A simple synthetic model for the electromagnetic fluctuating quantities can be obtained using a vector potential $\bv{A}$ that lies on the $(x,y)$ plane, namely $\bv{A} = (A_x(\bv{r},t),A_y(\bv{r},t),0)$. In the gauge where the scalar potential is zero, the electric field is given by $\bv{E} = (E_x,E_y,0) = - \partial \bv{A} /\partial t$, while the magnetic field lies in the direction perpendicular to the $(x,y)$ plane, namely $\bv{B} = (0,0,B) = \nabla \times \bv{A}$. With these assumptions the particle dynamics becomes bidimensional because the $z$ component of the Lorentz force is zero.

We consider a model where the electromagnetic fluctuations are due to the superposition of electric and magnetic fields generated by magnetised ``clouds'' which evolve in time
according to some prescribed laws. While in the original model by Fermi cosmic rays interact with
clouds only through usual collisions \cite{fermi,bouchet,NOI}, here we investigate the case where
the Lorenz force acts continuously on a test particle. We consider an $L \times L$ square as the
basic cell containing $N$ ``clouds''. We suppose that the basic
cell is repeated in space and that the two components of the vector potential
are thus constructed through the superposition
\begin{equation}
A_x = A_y = A_0 \sum_{n=1}^{N} \left[ \psi(\xi_n) +  \sum_{m} \psi(\tilde{\xi}_m) \right] \; ,
\label{eq-A}
\end{equation}
where $\psi = e^{-\xi_n}$, $\xi_n = |\bv{r} - \bv{r}_n(t)|/R$,
$\bv{r}_n(t) = (x_n(t),y_n(t))$ are the coordinates of the $n$--th cloud , $R$ represents the typical spatial extension of the potential produced by a single cloud, and $\tilde{\xi}_m = |\bv{r} - \tilde{\bv{r}}_m(t)|/R$, with $\tilde{\bv{r}}_m$ being the coordinates of the
obstacles in the neighbours of the basic cell. According to the hypothesis that the other
cells are duplicates of the basic cell, we have that 
$\tilde{\bv{r}}_m = \bv{r}_n + (iL,jL)$ where $i$ and $j$ are integer
numbers and at least one of them is non zero. Moreover, in order to simplify the sum in Eq. (\ref{eq-A}), $R$ and $N$ have been chosen in such a way that $R$ is typically smaller
than the distance between two nearby clouds, therefore in the summation $\sum_{m}$ of Eq. (\ref{eq-A}) it is sufficient to consider only the eight nearest neighours of the basic
cell (i.e. $(i,j) = (\pm 1,0), (0,\pm 1), (\pm 1, \pm 1), (\mp 1, \pm 1)$ ).
Since we are interested in investigating the diffusive properties of the model and not in reproducing the detailed properties of cosmic rays, at variance with the model by Fermi \cite{fermi} where clouds are allowed to move with a given bulk speed, here the motion of the obstacles has been chosen to be sinusoidal along both $x$ and $y$. That is, the coordinates of the $n$-th obstacle $x_n(t)$ and $y_n(t)$ are given by
\begin{eqnarray}
x_n(t) & = & x_{n0} + a_n \cos[\omega_n t + \alpha_n] \\
y_n(t) & = & y_{n0} + b_n \sin[\omega_n t + \beta_n] \; ,
\end{eqnarray}
where $x_{n0}$ and $y_{n0}$ are the initial coordinates of the $n$-th obstacle, $a_n$ and $b_n$ are the amplitudes of the oscillations of the $n$-th obstacle along $x$ and $y$ respectively, $\omega_n$ is the oscillation frequency, and $\alpha_n$ and $\beta_n$ are the initial oscillation phases, randomly chosen  within the interval $[0,2\pi]$. For the present paper we choose to keep the problem as simple as possible, so that we used the same values of motion amplitudes and frequencies for all clouds, namely $a_n = b_n = a$ and $\omega_n = \omega$.
The $N$ clouds contained in the basic cell are irregularly distributed, that is,
with random initial positions $(x_{n0},y_{n0})$.
%, that is,
%
%\begin{equation}
%R \leq \min\left\lbrace \frac{1}{2\sqrt{N}}-a_n , \frac{1}{2\sqrt{N}}-b_n\right\rbrace  \; .
%\end{equation}

The motion equations can be adimensionalised by using the following normalization factors: $B_0=A_0/L$ for magnetic field , $\omega_0=(q B_0)/m$ for frequency, $t_0=\omega_0 / 2 \pi$ for time, $v_0=L \omega_0$ for velocity, $E_0=A_0 \omega_0$ for electric field respectively, so that finally Eq. (\ref{eq-v}) can be
written as

%\begin{eqnarray}
%{dx \over dt} & = & v_x \label{eq-mot-x} \; ,\\
%{dy \over dt} & = & v_y \, ,
%\label{eq-mot-y}
%\end{eqnarray}
%
\begin{equation}
\frac{d \bv{v}}{d  t} =
\bv{E}(\bv{r},t) + \bv{v} \times \bv{B}(\bv{r},t) \; ,
%{d v_x \over dt} & = & v_y B + E_x \label{eq-mot-vx} \; ,\\
%{d v_y \over dt} & = & -v_x B + E_y \; .
\label{eq-mot-vy}
\end{equation}
where
\begin{equation}
B(\bv{r},t) = 
%{\partial A_y \over \partial x} - {\partial A_x \over \partial y} =
\sum_n \frac{1}{R} \frac{\partial \psi}{\partial \xi_n} \left\{
\frac{ [x-x_n(t)] - [y-y_n(t)] }{|\bv{r}-\bv{r}_n(t)|} \right\} + N.T.
\label{eq-B}
\end{equation}
\begin{equation}
E_i(\bv{r},t) = 
%- {\partial A_i \over \partial t} =
\sum_n \frac{1}{R} \frac{\partial \psi}{\partial \xi_n} \left\{
\frac{ [x-x_n(t)] dot{x}_n(t) + [y-y_n(t)] dot{y}_n(t) }{
|\bv{r}-\bv{r}_n(t)|} \right\} \bv{e}_i + N.T.
\label{eq-E}
\end{equation}
$\bv{e}_i$ denotes the unit vector
along the $i$--th direction on the plane, and $N.T.$ stands for the similar terms given
by the clouds in the nearest cells.

In the stationary limit case ($\omega=0$) the model is nothing but the electromagnetic reformulation
of the celebrated Lorentz model. The only (marginal) difference is the presence of a smooth potential.
For $\omega \neq 0$ one has a sort of Lorentz model where the obstacle positions are time dependent.
In Ref. \cite{papachristou} a dynamical system study in the case of few obstacles can be found.
%%%%%%%%%%%%%%%%%%%%%%%%%%%%%%%%%%%%%%%%%%%%%%%%%%%%%%%%%%%%%%%%%%%%%%%

\section{Test particle simulations}

\begin{figure*}
\begin{center}
\includegraphics[scale=0.45]{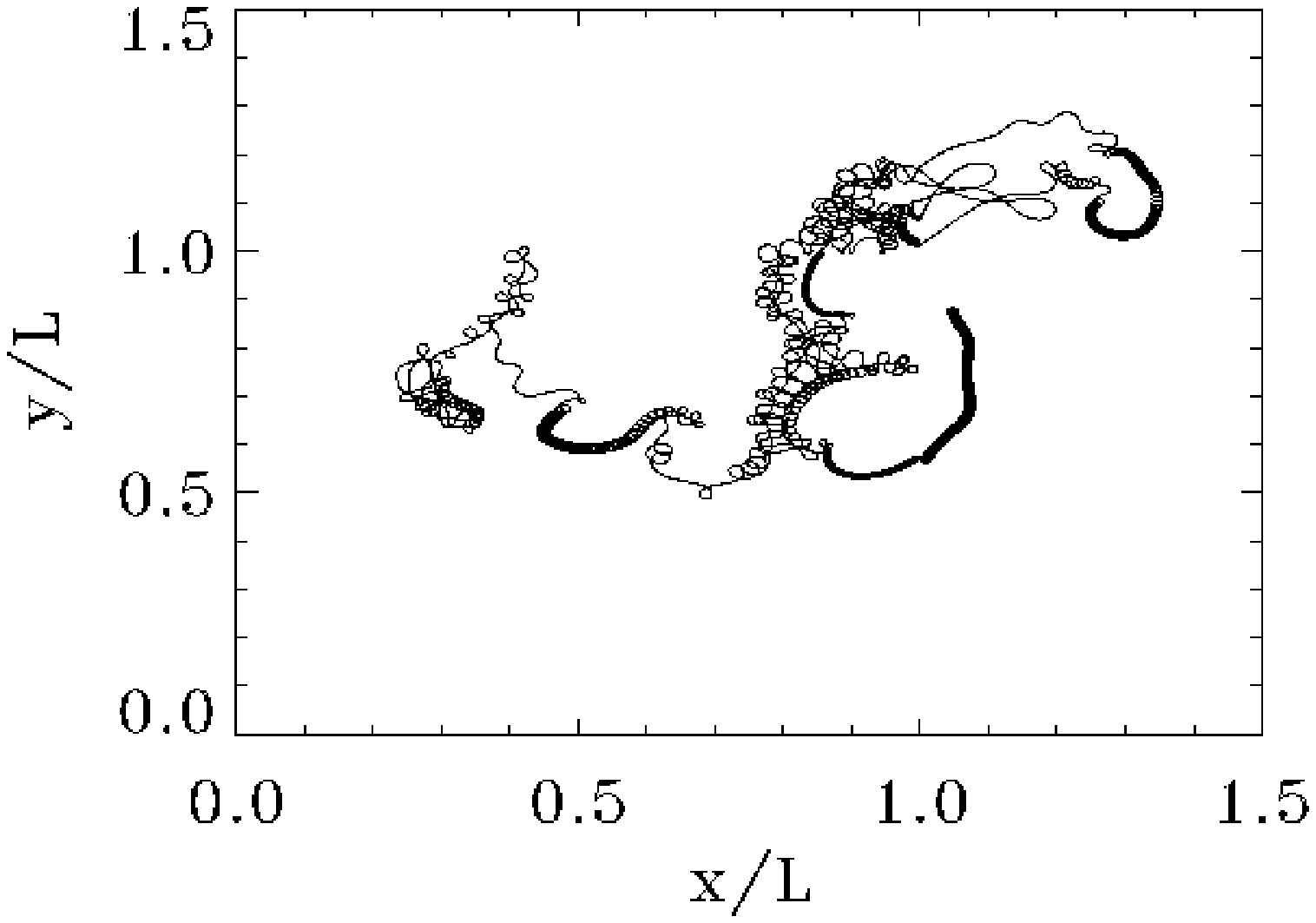}
\includegraphics[scale=0.45]{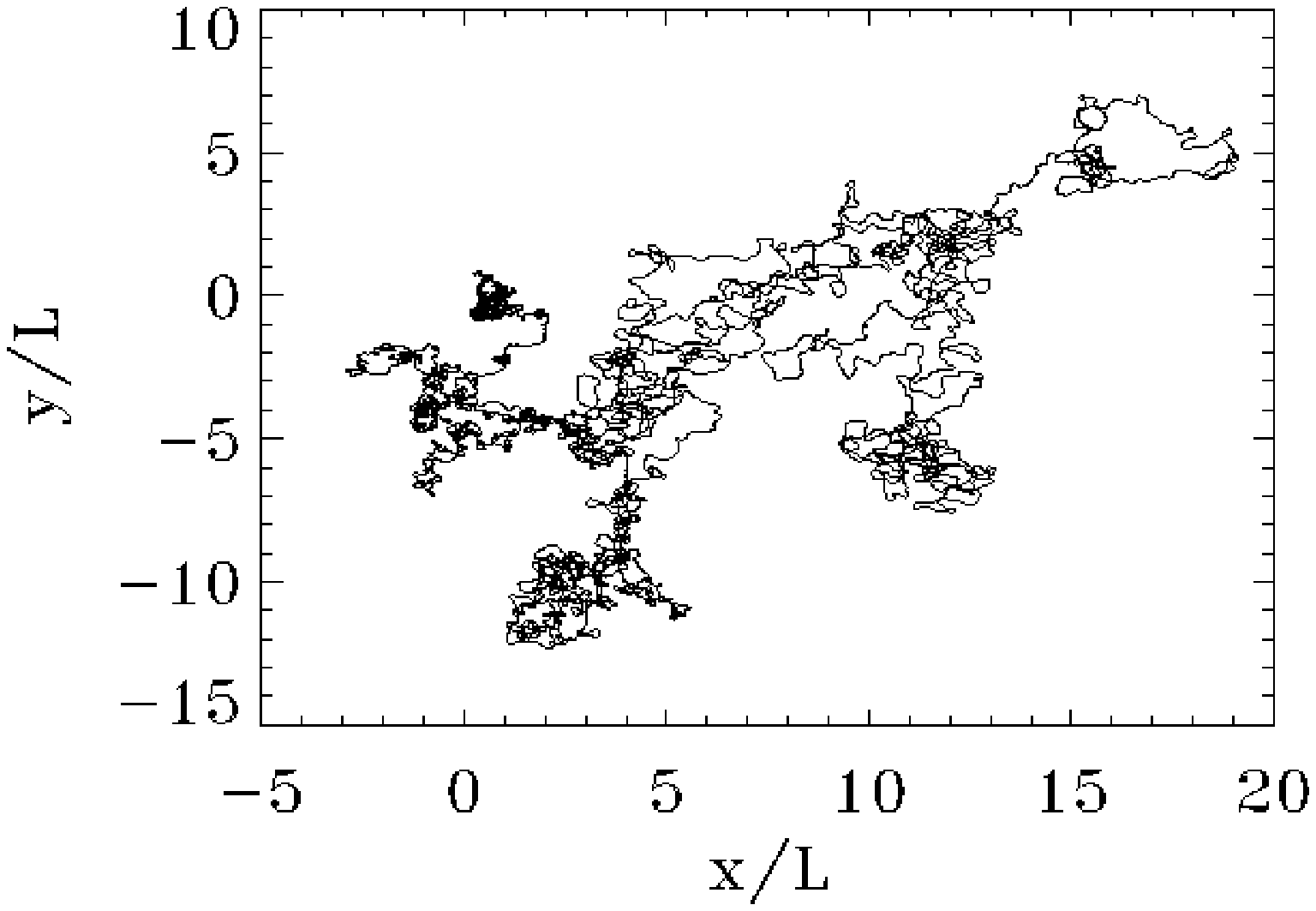}
\caption{Examples of particle trajectories for $\omega=0.01$ (left panel) and $\omega=10$ (right panel).}
\label{fig0}
\end{center}
\end{figure*}

The model described above is investigated through test particle numerical simulations.
The motion equations
%Eqs. %(\ref{eq-mot-x})-(\ref{eq-mot-vy})
are solved using a 4th order Runge-Kutta scheme. Particles are injected at random positions within the simulation box, with velocities extracted from a 2-D Maxwellian distribution $P(v_x,v_y) \propto \exp[-(v_x^2+v_y^2)/2v_{th}^2]$ with $v_{th} = 3 \times 10^{-2}$. Since we aim to investigate the diffusive properties of the system,
all particles have to be tracked for the same time interval. The results shown below are obtained by tracking $6 \times 10^3$ particles using the parameters $N=50$, $a = 10^{-3}$ and $R=0.07$, while three different values are used for the frequency, namely $\omega=0.01$, $\omega = 1$ and $\omega = 10$. We choose to vary only this last parameter, because preliminary results have clearly shown that the model is not significantly sensitive to variations of $a$ and $R$. In Fig. \ref{fig0} two examples of particles trajectories, for
$\omega=0.01$ and $\omega=10$ respectively, are shown.
%%%%%%%%%%%%%%%%%%%%%%%%%%%%%%%%%%%%%%%%%%%%%%%%%%%%%%%%%%%%%%%%%%%%%%%%%%%%%%%%%%%%%%%%%%%
%%%%%%%%%%%%%%%%%%%%%%%%%%%%%%%%%%%%%%%%%%%%%%%%%%%%%%%%%%%%%%%%%%%%%%%%%%%%%%%%%%%%%%%%%%%

\begin{figure}
\includegraphics[scale=0.50]{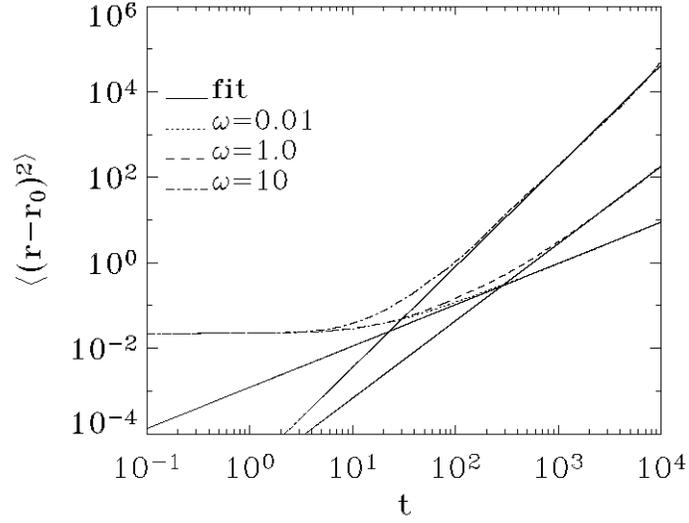}
\caption{Position mean square displacements for $\omega=0.01$ (dotted line), $\omega=1$ (dashed line), and $\omega=10$ (dash-dotted line). The power law best fits are shown as solid lines. The power law exponents obtained from these fits are given in Table \ref{tab1}.}
\label{fig1}
\end{figure}

\begin{figure}
\includegraphics[scale=0.50]{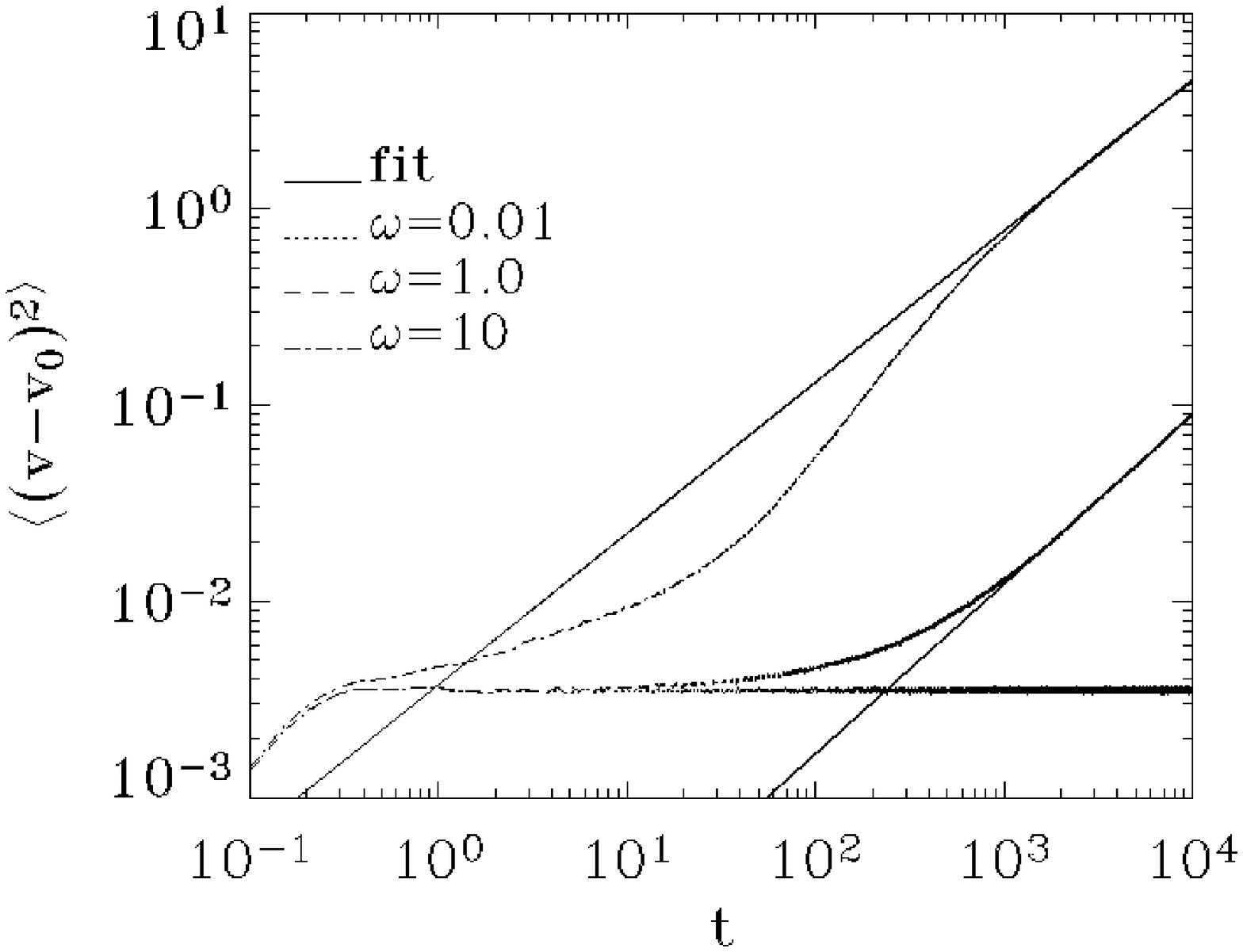}
\caption{Velocity mean square displacements for $\omega=0.01$ (dotted line), $\omega=1$ (dashed line), and $\omega=10$ (dash-dotted line). The power law best fits are shown as solid lines. The power law exponents obtained from these fits are given in Table \ref{tab1}.}
\label{fig1bis}
\end{figure}

In order to study the diffusive properties of the system in both position and velocity spaces,
the mean square position and velocity displacements are calculated as functions of time, that is,
$\langle \left[ \bv{r}(t) - \bv{r}_0 \right]^2 \rangle$ and $\langle \left[ \bv{v}(t) - \bv{v}_0 \right]^2 \rangle$, where brackets represent averages over the particle population. In the case of normal diffusion particles make a Brownian--like motion and $\langle \left[ \bv{r}(t) - \bv{r}_0 \right]^2 \rangle \sim t$. On the other hand, if a power law dependence of the mean square position
displacements on time is found for long times
\begin{equation}
\langle \left[ \bv{r}(t) - \bv{r}_0 \right]^2 \rangle \sim t^{2\nu_x} \; ,
\end{equation}
with $\nu_x \neq 1/2$, the diffusion is said to be anomalous, the cases
$\nu_x > 1/2$ and $\nu_x < 1/2$ being called superdiffusion and subdiffusion respectively. Diffusion in the velocity
space can be studied by investigating the scaling of the velocity mean square displacements
with time, that is, 
\begin{equation}
\langle \left[ \bv{v}(t) - \bv{v}_0 \right]^2 \rangle \sim t^{2\nu_v} \; .
\end{equation}
The mean square displacements of position and velocity are shown in Fig. \ref{fig1} and Fig. \ref{fig1bis} respectively for the three vales of $\omega$ which are used. The best fits with power laws are also shown as solid lines. 
%%%%%%%%%%%%%%%%%%%%%%%%%%%%%%%%%%%%%%%%%%%%%%%%%%%%%%%%%%%%%%%%%%%%%%%%%%%%%%%%%%%%%%%%%%%
%%%%%%%%%%%%%%%%%%%%%%%%%%%%%%%%%%%%%%%%%%%%%%%%%%%%%%%%%%%%%%%%%%%%%%%%%%%%%%%%%%%%%%%%%%%
%%%%%%%%%%%%%%%%%%%%%%%%%%%%%%%%%%%%%%%%%%%%%%%%%%%%%%%%%%%%%%%%%%%%%%%%%%%%%%%%%%%%%%%%%%%
%%%%%%%%%%%%%%%%%%%%%%%%%%%%%%%%%%%%%%%%%%%%%%%%%%%%%%%%%%%%%%%%%%%%%%%%%%%%%%%%%%%%%%%
The values of the $\nu_x$ and $\nu_v$ exponents obtained from the fits are given in Table \ref{tab1}.
\begin{table}
\caption{Power law exponents $\nu_x$ and $\nu_v$ obtained the from best fits of
position and velocity mean square displacements, respectively.}
\label{tab1}
\begin{center}
\begin{tabular}{|c|c|c|c|}
\hline
         & $\omega=0.01$ & $\omega=1$ &  $\omega=10$ \\
\hline
$\nu_x$ &$0.49$ & $0.89$ & $1.19$ \\
$\nu_v$  & $4.7 \times 10^{-4}$ &$0.42$  &$0.39$ \\
\hline
\end{tabular}
\end{center}
\end{table}
For $\omega=0.01$ the diffusion is nearly normal, since $\nu_x \simeq 1/2$ and the velocity mean square displacement remains constant after an initial transient. Such an exponent value indicates that the system
is in a standard diffusion regime. For both $\omega=1$ and $\omega=10$ a superdiffusive behavior in position is observed, and the system seems to exhibit diffusion also in velocity, with the velocity
mean square displacements showing a power law dependence. As a consequence of the time correlations, the relation between scaling exponents $\nu_x$ and $\nu_v$ is not simple \cite{stawicki,bouchet}. In fact, a straightforward calculation shows that 
\begin{equation}
\langle \left[ \bv{r}(t) - \bv{r}_0 \right]^2 \rangle \simeq 2 \int_0^t \int_0^{t-t'} \langle |\bv{v}(t')|^2\rangle C(t',\tau) \, d t' d \tau
\end{equation}
where $C(t',\tau) = \langle \bv{v}(t')\bv{v}(t'+\tau)\rangle/\langle |\bv{v}(t')|^2\rangle$. In the classical diffusion process, where acceleration is absent, $\langle |\bv{v}(t')|^2\rangle$ is constant,  $C(t',\tau)$ is independent of $t'$ and $\langle \left[ \bv{r}(t) - \bv{r}_0 \right]^2 \rangle$
becomes proportional to $t$ by integrating over $t'$. Therefore anomalous diffusion in the position space is entirely due to long time tails of $C(\tau)$. However, in the absence of stationarity the presence of $C(t',\tau)$ makes the scaling exponents $\nu_x$ and $\nu_v$ not related in a simple way.
In this situation only the upper bound $\nu_x \leq \nu_v + 1$ can be derived \cite{bouchet}. When the electric field is strong, or in other words when very strong time correlations of velocities at different times are present, the equality $\nu_x = \nu_v + 1$ (i.e. the maximum allowed difference between $\nu_x$ and $\nu_v$) seems to occur \cite{bouchet,stawicki}. This indicates that the precise relationship between scaling exponents in the position and velocity spaces is strongly dependent on the correlations introduced in the model. Different models are characterised by different scaling properties. For example in the model of Vlahos et al. \cite{vlahos}, where different correlations are used,
$\nu_x \simeq 1.15$ is found for position space diffusion, which is similar to the
value obtained in the present model for $\omega=10$, but velocity diffusion is weaker, namely $\nu_v \simeq 0.17$ with respect $\nu_v \simeq 0.39$ (cf. Table \ref{tab1}).
%%%%%%%%%%%%%%%%%%%%%%%%%%%%%%%%%%%%%%%%%%%%%%%%%%%%%%%%%%%%%%%%%%%%%%%%%%%%%%%%%%%%
%\begin{table} [h]
%\begin{center}
%\begin{tabular}{|r||r|r||r|r||r|r|} \hline
%  &\multicolumn{2}{c|}{$\omega=0.01$} &\multicolumn{2}{c|}{$\omega=1$}&\multicolumn{2}{c|}{$\omega=10$}\\
%\cline{2-7}
%  &$P(v_{x})$ & $P(v_{y})$ &$P(v_{x})$ & $P(v_{y})$ & $P(v_{x})$ & $P(v_{y})$\\
%\hline
%\hline
%mean & %$4\times10^{-4}$&$4\times10^{-4}$&$7\times10^{-4}$&$2\times10^{-3}$&$2.1\times10^{-2}$&$1.1\times10^{-3}$\\
%standard deviation   &$3\times10^{-2}$& $3\times10^{-2}$& $2.1\times10^{-1}$& %$2.1\times10^{-1}$& 1.49&1.50  \\
%skewness %&$9.6\times10^{-3}$&$1.1\times10^{-2}$&$-3.1\times10^{-2}$&$3.4\times10^{-2}$& %$-8.8\times10^{-3}$& $1.6\times10^{-2}$  \\
%kurtosis & %$-9.8\times10^{-2}$&$-3.1\times10^{-2}$&$-2.8\times10^{-1}$&$-2.2\times10^{-1}$&$-5\ti%mes10^{-1}$&$-4.5\times10^{-1}$   \\
%\hline
%\end{tabular}
%\end{center}
%\end{table}
%%%%%%%%%%%%%%%%%%%%%%%%%%%%%%%%%%%%%%%%%%%%%%%%%%%%%%%%%%%%%%%%%%%%%%%%%%%%%%%%%%%%
%\begin{table} [h]
%\begin{center}
%\begin{tabular}{|r|r|r|} \hline
%          & $P(v_{x})$   & $P(v_{y})$ \\
%\hline
%\hline
%mean  &$-3\times10^{-4}$ &$-1\times10^{-4}$\\
%standard deviation & $3\times10^{-2}$ & $3\times10^{-2}$ \\
%skewness &  $2.3\times10^{-2}$ &$5.4\times10^{-2}$ \\
%kurtosis & $8.1\times10^{-2}$ & $-8\times10^{-2}$ \\
%\hline
%\end{tabular}
%\end{center}
%\end{table}
%%%%%%%%%%%%%%%%%%%%%%%%%%%%%%%%%%%%%%%%%%%%%%%%%%%%%%%%%%%%%%%%%%%%%%%%%%%%%%%%%%%%

\begin{figure*}
\includegraphics[scale=0.60]{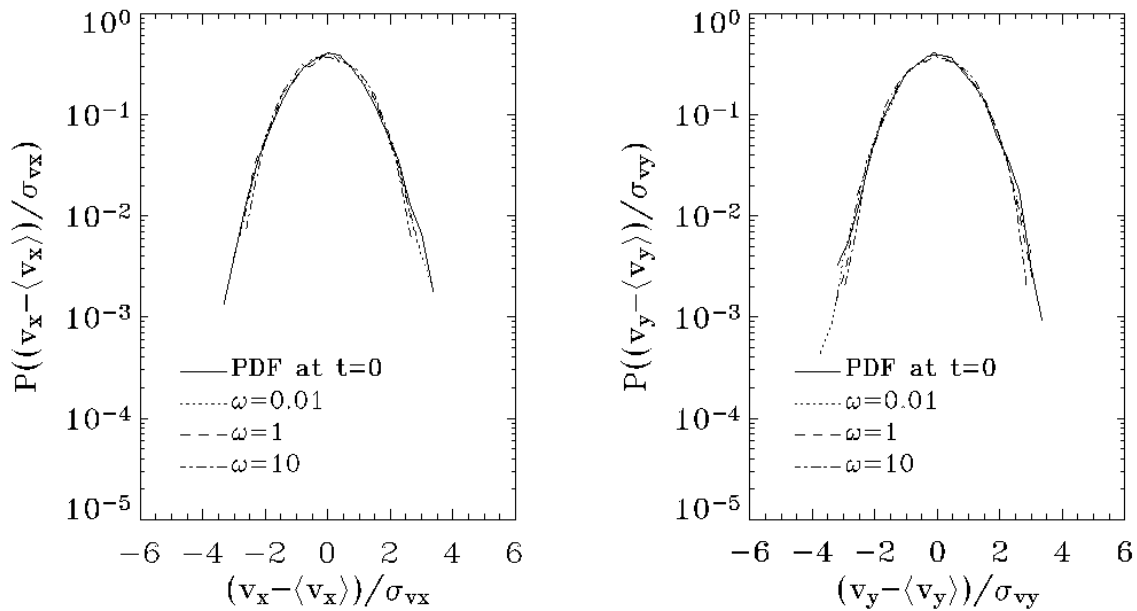}
\caption{PDFs of the standardised velocity components at $\omega=0.01$ (dotted lines), $\omega=1$ (dashed lines), and $\omega=10$  (dash-dotted lines), collected at the end of the numerical simulations. The PDFs at $t=0$ are shown as solid lines.}
\label{fig2}
\end{figure*}

The probability density functions (PDFs) of the two velocity components
at different $\omega$, collected at the end of the numerical simulations, i.e. $t=10^4$,
are shown in Fig. \ref{fig2} together with the velocity PDFs constructed at $t=0$,
which are the same in all cases. 
%%%%%%%%%%%%%%%%%%%%%%%%%%%%%%%%%%%%%%%%%%%%%%%%%%%%%%%%%%%%%%%%%%%%%%%%%%%%%%%%%%%%%%%%%%%
%%%%%%%%%%%%%%%%%%%%%%%%%%%%%%%%%%%%%%%%%%%%%%%%%%%%%%%%%%%%%%%%%%%%%%%%%%%%%%%%%%%%%%%%%%%
In order to compare the shape of the PDFs for different $\omega$, the PDFs are calculated after transforming the variables $v_x$ and $v_y$ into their standardised form by subtracting their averages $\langle v_x \rangle$ and $\langle v_y \rangle$ respectively and dividing by the standard deviations $\sigma_{v_x}$
and $\sigma_{v_y}$ (after this transformation the variables have zero mean and unit standard
deviation). The shape of the velocity PDFs does not change significantly with respect to the initial one, remaining nearly Gaussian. However, the standard deviations of $v_x$ and $v_y$ increase as $\omega$ increases (see Table \ref{tab2}), i.e. the velocity distributions
become broader.
%%%%%%%%%%%%%%%%%%%%%%%%%%%%%%%%%%%%%%%%%%%%%%%%%%%%%%%%%%%%%%%%%%%%%%%%%%%%%%%%%%%%%
\begin{table}[!ht]
\caption{Standard deviations of the two velocity components collected at the
end of the numerical simulations for $\omega=0.01$, $\omega=1$, and $\omega=10$.
As a comparison the standard deviations
of the initial velocities are shown in the second column.}
\label{tab2}
\begin{center}
\begin{tabular}{|c|c|c|c|c|}
\hline
& Initial & $\omega=0.01$ & $\omega=1$ & $\omega=10$ \\
\hline
$\sigma_{v_x}$ & 0.03 & 0.03 & 0.21 & 1.49 \\
\hline
$\sigma_{v_y}$ & 0.03 & 0.03 & 0.21 & 1.49 \\
\hline
\end{tabular}
\end{center}
\end{table}
%%%%%%%%%%%%%%%%%%%%%%%%%%%%%%%%%%%%%%%%%%%%%%%%%%%%%%%%%%%%%%%%%%%%%%%%%%%%%%%%%%%%%

\section{Conclusions}

In conclusion we investigated a model of particle diffusion and acceleration in stochastic electromagnetic fields, focusing on the diffusive properties in both position and velocity spaces. As the oscillation
frequency of the magnetised ``clouds'', which produce the electromagnetic fields, increases, the diffusion
in position space goes from a Brownian-like regime to anomalous diffusion, with ballistic motion for
the highest investigated frequency. Correspondingly, diffusion in velocity space also shows a change
from Brownian-like behavior to anomalous diffusion. However the relation between the diffusion exponents in position and velocity does not seem to be simple, depending on the detailed properties of the time correlation
introduced in the model at hand. In the cases where anomalous diffusion is present, a broadening of the
particle velocity distribution is also found, even if their shape remains nearly Gaussian. In other words,
the anomalous diffusion mechanism at work is associated to an energization of the bulk particle population
and not to the existence of high energy tails in the velocity distribution.

\acknowledgments
Numerical calculations have been performed in the framework of HPCC (Center of Excellence for High Performance Computing) of the University of Calabria.

%\bibliography{/home/lepreti/latex/fabio}

\end{document}